\documentstyle[prl,aps,epsfig]{revtex}

\widetext

\begin{document}

\title{\bf Stochastic Phase Space Localisation \\ 
for a Single Trapped Particle}

\author{Stefano Mancini\footnote{Present address:
Dipartimento di Fisica, Universit\`a di Milano,
Via Celoria 16, I-20133 Milano, Italy.}, 
David Vitali and Paolo Tombesi}

\address{Dipartimento di Matematica e Fisica, Universit\`a di
Camerino, via Madonna delle Carceri I-62032 Camerino \\
and Istituto Nazionale per la Fisica della Materia, Camerino, Italy\\}

\date{\today}

\maketitle

\begin{abstract}
We propose a feedback scheme to control the vibrational 
motion of a single trapped particle 
based on indirect measurements of its position.
It results the possibility of a motional phase space 
uncertainty contraction, 
corresponding to cool the particle close to the motional
ground state.
\end{abstract}


\section{Introduction}

In recent years there has been an increasing interest on trapping 
phenomena
and related cooling techniques  \cite{pg}.
Some years ago it has been shown that, using resolved sideband
cooling, a single ion can be trapped 
and cooled
down near to its zero-point vibrational energy state \cite{wineprl}
and recently, analogous results have been obtained for neutral atoms 
in optical lattices \cite{optlat}. 
The possibility to control trapped particles, indeed,
gave rise to new models
in quantum computation \cite{ciraczoller}, in which information is encoded
in two internal electronic states of the ions and the
two lowest Fock states of a vibrational collective mode
are used to transfer and manipulate quantum information between them. 
It may happen however, that a trapped ion that is a favorable candidate
for quantum information processing since it posseses a hyperfine
structure with long coherence times (as for example
$^{25}$Mg$^+$ \cite{peik}), is not suitable for resolved sideband 
cooling. In such cases it may be helpful to have an alternative
cooling technique, which can be applied when resolved sideband 
cooling is impractical to use.

In this paper we present a
way to control the motion of a trapped particle, which is able to
give a significant phase-space-localisation. 
The basic idea of the scheme is to realize
an effective and continuous measurement of
the position of the trapped particle and then apply a feedback
loop able to decrease the position fluctuations. 
Due to the continuous nature of the measurement and to the
effect of the trapping potential coupling the particle
position with its momentum, feedback will realize an effective
phase-space localisation. 

With this respect there are some analogies between the present method
and resolved-sideband stimulated Raman cooling \cite{monroe},
which can be viewed as a sort of feedback scheme.
In fact, one of the two Raman lasers performs an effective 
measurement of the vibrational number by changing the particle internal
state only if it is an excited vibrational state. The second
Raman laser performs instead the feedback step, because
it puts the particle back in the initial internal state, after having
removed a vibrational quantum. The feedback scheme proposed here
measures the particle position rather than its energy and tries to achieve
cooling as phase-space localisation, using the particle 
oscillatory motion to mix position and momentum quadratures.

A second analogy is given by the fact that
the proposed method needs a Doppler pre-cooling stage,
as it happens for resolved sideband cooling. In fact,
the effective trapped particle position measurement
is realized only in the Lamb-Dicke regime, i.e. when the recoil 
energy is much
smaller than the energy of a vibrational quantum, which
can be obtained only when the particle has undergone a preliminary
cooling stage. Our scheme will
provide therefore further phase space localisation and cooling.

The paper is organized as follows. In section II we show
how to realize the indirect continuous measurement of the position
by coupling the trapped particle with a standing wave. In section III
we shall introduce the feedback loop, in section IV we shall
study the properties of the stationary state in the presence
of feedback and section V is for concluding remarks.

\section{Continuous position measurement}

We consider a generic particle trapped in an effective harmonic potential.
For simplicity we shall consider the one-dimensional
case, even if the method can be in principle generalized to the
three-dimensional case. This particle can be an ion trapped by a linear
rf-trap \cite{nist} or a neutral atom in an optical trap \cite{optlat,sara}. 
Our scheme however does not depend on the specific trapping
method employed and therefore we shall always refer from now on
to a generic trapped ``atom''.

The trapped atom of mass $m$, 
oscillating with frequency $\nu$ along the
$\hat{x}$ direction and with position
operator $x=x_{0}(a+a^{\dagger})$, $x_{0}=(\hbar/2m\nu)^{1/2}$,
is coupled to a standing wave with frequency $\omega_{b}$,
wave-vector $k$ along $\hat{x}$ and
annihilation operator $b$. The standing wave is quasi-resonant with the 
transition between two internal atomic levels $|+\rangle$ and $|-\rangle $.
The Hamiltonian of the system is \cite{qo}
\begin{equation}
H=\frac{\hbar \omega_{0}}{2}\sigma_{z} +\hbar\nu a^{\dag}a+
\hbar\omega_{b} b^{\dagger}b +i\hbar \epsilon (\sigma_+ +\sigma_-)
(b-b^{\dagger}) \sin\left(kx+\phi\right) \,,
\end{equation}
where $\sigma_z= |+\rangle \langle +|-|-\rangle \langle -|$, 
$\sigma_{\pm}=|\pm \rangle \langle \mp |$, and $\epsilon$
is the coupling constant.
In the interaction representation with respect to
$H_0=\hbar \omega \left(b^{\dagger} b +\frac{\sigma_{z}}{2}\right)$,
where $\omega \sim \omega_b$ will be specified later, and making the rotating
wave approximation, this Hamiltonian becomes
\begin{equation}
H=\frac{\hbar \Delta}{2}\sigma_{z} +\hbar\nu a^{\dag}a+
\hbar \delta b^{\dagger}b +i\hbar \epsilon (\sigma_+ b - \sigma_-b^{\dagger})
 \sin\left(kx+\phi\right) \,, \label{hrwa}
\end{equation}  
where $\Delta=\omega_0-\omega$ and $\delta = \omega_b-\omega$ are the
atomic and field mode detuning, respectively.

There are now two different ways for realizing an effective
continuous measurement of the atom position and we shall describe them
separately, even if they present many similarities.

\subsection{The Resonant case}\label{sub1}

We consider the case when the standing wave is perfectly resonant
with the $|+\rangle \leftrightarrow |-\rangle $
transition, i.e., $\omega_b= \omega_0$. It is therefore convenient to choose
the frequency of the rotating frame $\omega = \omega_b= \omega_0$ in this case, so that both detunings
are equal to zero. Moreover we shall consider the case 
of a very intense standing wave, so that it can be treated classically, that is,
$b$ can be replaced by the c-number $\beta$. Choosing the phase of the field such that
$\beta = -i |\beta |$, Hamiltonian (\ref{hrwa}) becomes
\begin{equation}
H=\hbar\nu a^{\dag}a+
\hbar \epsilon |\beta |\sigma_x 
 \sin\left(kx+\phi\right) \,, \label{hris1}
\end{equation}
where $\sigma_x = \sigma_+ + \sigma_-$.
If we finally set the spatial phase $\phi =0$ (i.e. the atom is trapped near  
a node of the classical standing wave) and assume the Lamb-Dicke regime, we can 
approximate the sine term at first order and get
\cite{danrisk,cirac}
\begin{equation}\label{Hini}
H=\hbar\nu a^{\dag}a+\hbar\chi\sigma_xX\,,
\end{equation}
where 
$\chi = 2\epsilon |\beta |kx_0$ is the effective
coupling constant between the internal and the 
vibrational degrees of freedom,
and $X=(a+a^{\dag})/2$ is the dimensionless position operator of the trapped
atom. This Hamiltonian shows how one can realize an effective 
measurement of the atomic position. In fact,
the atom displacement away from the electric field node
increases the probability of electronic excitation, and 
hence displacements can be 
monitored by means of the atomic fluorescence.
Therefore, the two-level (sub)system can be used as a meter to 
measure the position quadrature $X$.

The evolution 
equation for the total density operator $D$ for the vibrational degree of
freedom and the internal states is determined by Hamiltonian (\ref{Hini})
and by the terms describing the spontaneous emission
from the level $|+\rangle $ responsible for the fluorescence, 
\begin{equation}
{\cal L}_{spont}D =
\frac{\kappa}{2}
\left(2\sigma_-D\sigma_+-\sigma_+\sigma_-D
-D\sigma_+\sigma_-\right)\;,
\label{spont}
\end{equation}
where $\kappa$ is the spontaneous emission rate.  Here 
we have neglected the recoil and the associated heating
of the vibrational motion. This is reasonable
in the Lamb-Dicke limit we have assumed from the beginning, since
the associated heating rate is given by $\kappa (kx_0)^2$
vibrational quanta per second, which is
negligible for a sufficiently small Lamb-Dicke parameter $kx_0$.
In practical situations, also other heating mechanisms
exist, caused by technical imperfections such as 
the fluctuations of trap parameters due to ambient fluctuating electrical
fields
in the ion trap case \cite{nist}, and due to laser intensity noise
and beam-pointing fluctuations in the case of far-off resonance optical traps 
(see Ref.~\cite{sara} and references therein).
We assume the presence of this heating due to trap imperfections, 
and we describe it with the following term in the master equation, 
characterized by a heating rate $\gamma_h$ (see \cite{sara2})
\begin{equation}
{\cal L}_{h}D=\frac{\gamma_h}{2}
\left(2a D a^{\dag}-a^{\dag}aD-D a^{\dag}a\right)
+\frac{\gamma_h}{2}
\left(2a^{\dag}D a-aa^{\dag}D-D aa^{\dag}\right)\,.
\label{liu}
\end{equation}
The heating rate $\gamma_h$ has not to be too large, 
in order to stay within the assumed
Lamb-Dicke regime. The Lamb-Dicke condition also implies that
the trapped atom has to be initially prepared in a sufficently cold state,
i.e., an effective thermal state with, say, a mean vibrational number 
$n_0 \sim 10$. This can be obtained with a preliminary Doppler cooling stage,
which is then turned off at $t=0$ and replaced by the proposed feedback
cooling scheme. We shall see that
our scheme is able to further cool the trapped atom, close to the 
ground state, even in the presence of moderate heating processes.

The resulting master equation for the internal and vibrational
degrees of freedom is
\begin{equation}
\dot{D}={\cal L}_{h}D -\frac{i}{\hbar}\left[H,D\right]
+\frac{\kappa}{2}
\left(2\sigma_-D\sigma_+-\sigma_+\sigma_-D
-D\sigma_+\sigma_-\right)\;.
\label{evoluzero}
\end{equation}
Let us now see how to realize the
continuous position measurement.
It has been recently shown that when excited by a low intensity 
laser field, a single trapped atom emits its fluorescent light mainly
within a quasi-monochromatic elastic peak \cite{hof}.
The fluorescent light spectrum was measured by heterodyne detection. 
By improving the technique it does not seem impractical to get a 
homodyne detection of the single-ion fluorescent light.
In Ref. \cite{vog}, it was shown how one could achieve such a measurement.
Thus, by exploiting the resonance fluorescence it could be possible to 
measure the quantity
$\Sigma_{\varphi }=\left(\sigma_-e^{-i\varphi }
+\sigma_+e^{i\varphi }\right)$ through homodyne detection of 
the field 
scattered by the atom along a certain direction \cite{qo}.
In fact, the detected field may be written in terms of the 
dipole moment operator 
for the transition
$|-\rangle\leftrightarrow|+\rangle$ as \cite{qo}
\begin{equation}\label{Escat}
E^{(+)}_s(t)=\sqrt{\eta\kappa}\sigma_-(t)\,,
\end{equation}
where $\eta$ is an overall quantum efficiency accounting 
for the detector efficiency and the fact that only a small
fraction of the fluorescent light is collected and superimposed
with a mode-matched oscillator.

As a consequence of (\ref{Escat}), the homodyne photocurrent 
will be \cite{quadra}
\begin{equation}\label{photoc}
I(t)=2\eta\kappa\langle\Sigma_{\varphi}(t)\rangle_c
+\sqrt{\eta\kappa}\xi(t)\,,
\end{equation}
where the phase
$\varphi$ is related to the local oscillator, which, since we have
assumed the resonance condition $\Delta =0$, in the
present case is provided by the same driving field generating
the classical standing wave. The subscript
$c$ in Eq.~(\ref{photoc}) 
denotes the fact that the average is performed on the state 
conditioned on
the results of the previous measurements and
$\xi(t)$ is a Gaussian white noise
\cite{quadra}.
In fact, the continuous monitoring of the electronic mode
performed through the homodyne measurement,
modifies the time evolution of the whole system, and the 
state conditioned on the result of
measurement, described by a stochastic conditioned density matrix 
$D_c$, evolves
according to the following stochastic differential equation 
(considered in the Ito sense)
\begin{eqnarray}\label{Dceq}
{\dot D}_c&=&{\cal L}_{h}D_c-\frac{i}{\hbar}
\left[H,D_c\right] +\frac{\kappa}{2}\left(2\sigma_-D_c\sigma_+-\sigma_+\sigma_
-D_c-D_c\sigma_+\sigma_-\right)
\nonumber\\
&+&\sqrt{\eta\kappa}\,\xi(t)\left(
e^{-i\varphi}\sigma_-D_c+e^{i\varphi}D_c\sigma_+
-2\langle\Sigma_{\varphi}\rangle_cD_c\right)\,. \nonumber 
\end{eqnarray}

Since we are considering a strong fluorescent transition
it is reasonable to assume that the spontaneous emission rate
$\kappa$ is large, i.e.,
$\kappa \gg \chi$. This means that the internal two-level
system is heavily damped and that it 
will almost always  be 
in its lower state $|-\rangle $. 
This allows us to adiabatically eliminate the
internal degree of freedom  
and to perform a perturbative calculation in the small 
parameter $\chi /\kappa$, obtaining (see also Ref.~\cite{TV})  
the following expansion for the total conditioned density matrix $D_c$           
\begin{equation}\label{Dofrho}
D_c=\rho_c\otimes|-\rangle\langle-|
-i\frac{\chi}{\kappa}\left(
X\rho_c\otimes|+\rangle\langle-|
-\rho_c\otimes|-\rangle\langle+|X\right)\,,
\end{equation}
where $\rho_c={\rm Tr}_{el}\,D_c$ is the reduced conditioned
density matrix for 
the vibrational motion. In the adiabatic regime, the internal dynamics
instantaneously follows the vibrational one and therefore
one gets 
information 
on the position dynamics $X$ by observing the quantity $\Sigma_{\varphi}$. 
The relationship between the conditioned mean values follows 
from Eq.(\ref{Dofrho})
\begin{equation}\label{homoflo}
\langle\Sigma_{\varphi}(t)\rangle_c
=\frac{\chi}{\kappa}\langle X(t)\rangle_c \sin\varphi\,.
\end{equation}

Moreover, if we adopt the perturbative expression (\ref{Dofrho}) for 
$D_c$ in (\ref{Dceq}) and perform
the trace over the internal mode, we get an equation for the 
reduced density matrix $\rho_c$
conditioned to the result of the measurement of the observable 
$\langle \Sigma_{\varphi}(t)\rangle _c$, and therefore 
$\langle X(t)\rangle _c$
\begin{eqnarray}\label{rhoceq}
{\dot\rho}_c&=&{\cal L}_{h}\rho_c-i\nu \left[a^{\dagger}a,\rho_c\right]
-\frac{\chi^2}{2\kappa}
\left[X,\left[X,\rho_c\right]\right]\nonumber\\
&+&\sqrt{\eta\chi^2/\kappa}\,\xi(t)
\left(ie^{i\varphi}\rho_cX-ie^{-i\varphi}X\rho_c
+2\sin\varphi\langle X(t)\rangle_c\rho_c\right)\,.
\end{eqnarray}
This equation describes the stochastic evolution of the vibrational state
of the trapped atom conditioned to the result of the continuous
homodyne measurement of the resonance fluorescence. The double commutator
with $X$ is typical of quantum non-demolition (QND)
measurements of the position. However this indirect measurement is not
properly QND because of the presence of the vibrational bare Hamiltonian
$\hbar \nu a^{\dagger} a$ mixing the position quadrature with the
momentum.

\subsection{The Off-Resonant Interaction}\label{sub2}

One can realize an effective indirect measurement of the atomic
position also in the opposite limit of large detuning
between the standing wave field and the internal transition.
In fact, when the internal detuning $\Delta $ is very large
($\Delta \gg \nu, \delta, \kappa, \epsilon$) the excited level
$|+\rangle $ can be adiabatically eliminated:
the state of the whole system (atom+standing wave mode) can be written as
$\psi_+ |+\rangle + \psi_- |-\rangle $, and adding a constant term 
$\hbar \Delta/2$ to the Hamiltonian (\ref{hrwa}),
the corresponding Schr\"odinger
equations will be
\begin{eqnarray}
i \dot {\psi}_+ &=& (\Delta +\nu a^{\dagger} a+ \delta b^{\dagger} b)
\psi_+ +i \epsilon \sin(kx+\phi) b \psi_- \label{adiab1} \\ 
i \dot {\psi}_- &=& (\nu a^{\dagger} a+ \delta b^{\dagger} b)
\psi_- -i \epsilon \sin(kx+\phi) b^{\dagger} \psi_+ \;.  
\label{adiab2}
\end{eqnarray}
In the adiabatic limit of very large $\Delta$ we can 
neglect the time derivative in (\ref{adiab1}) and put
\begin{equation}
\psi_+ \simeq -i \frac{\epsilon}{\Delta} \sin(kx+\phi) b \psi_- \;.
\end{equation}
Inserting this equation into (\ref{adiab2}), 
one gets an equation for $\psi_-$ which is equivalent 
to have the following effective Hamiltonian for the vibrational
motion of the atom and the standing wave mode alone
\begin{equation}\label{H2}
H=\hbar \delta b^{\dagger}b
+\hbar\nu a^{\dagger}a
-\hbar\frac{\epsilon^2}{\Delta} b^{\dag}b\sin^2
\left(kx+\phi\right)\; .
\end{equation}
If we now set the spatial phase $\phi=\pi/4$, we can rewrite
(\ref{H2}) as
\begin{equation}\label{H2b}
H=\hbar \left(\delta -\frac{\epsilon ^2}{2 \Delta}\right)
b^{\dagger}b
+\hbar\nu a^{\dagger}a
-\hbar\frac{\epsilon^2}{2\Delta} b^{\dag}b\sin 
\left(2kx\right)\; .
\end{equation}
It is clear that in this case it is convenient to choose
the frequency $\omega$ of the rotating frame so that
$\delta = \epsilon^2 /2\Delta$. The Hamiltonian (\ref{H2b}) 
assumes the desired form 
when the Lamb-Dicke regime is again assumed so to approximate the sine term
with its argument, and when the case of an intense standing wave is considered.
However, in this case we shall not neglect the quantum fluctuations
of the standing wave field, and we shall 
make the replacement $b \to \beta + b$,
where $\beta \gg 1$ describes the classical coherent steady state of the 
radiation mode and $b$ is now the annihilation operator describing
the quantum fluctuations. One gets
\begin{equation}\label{H2c}
H=\hbar\nu a^{\dagger}a
-\hbar\frac{\epsilon^2}{\Delta} kx (|\beta|^2+\beta^* b+\beta b^{\dagger})\; .
\end{equation}
Shifting the origin along the $x$ direction
by the quantity $ \hbar \epsilon^2 k |\beta |^2 /\Delta m \nu^2 $,  
one finally gets an effective Hamiltonian analogous to
that of the resonant case (\ref{Hini})
\begin{equation}\label{Heff1}
H=\hbar\nu a^{\dag}a
+\hbar\chi YX\,,
\end{equation}
where now $\chi=-4|\beta | k x_0 \epsilon^2/\Delta$ and the
atomic polarization $\sigma_x$ is replaced by the standing wave
field quadrature
$Y=(be^{-i\phi_{\beta}}+b^{\dag}e^{i\phi_{\beta}})/2$,
where $\phi_{\beta}$ is the phase of the classical amplitude $\beta$.
This means that in this case the ``meter'' is represented by the cavity mode, and
that an effective continuous measurement of the position
of the trapped atom is provided by
the homodyne measurement of the light outgoing from the cavity.
This measurement allows in fact to obtain 
the quantity $Y_{\varphi}=(ae^{-i\varphi}+a^{\dag}e^{i\varphi})/2$, which
is analogous to the quantity $\Sigma_{\varphi}$ of the previous
Section. Therefore, all the steps leading to Eq. (\ref{rhoceq}) in the
preceding subsection, can be 
repeated here, with the appropriate changes. In this non-resonant
case, $D$ now refers to the density matrix of the system composed by
vibrational mode and the standing wave mode and the spontaneous emission 
term in the master equation (\ref{evoluzero}) has to be replaced by the 
formally analogous term describing damping of the standing wave mode due to
photon leakage. This is equivalent to interpret the parameter $\kappa$
as a cavity mode decay rate in this case and to replace $\sigma_-$ with $b$,
$\sigma_+$ with $b^{\dagger}$, and $\Sigma_{\varphi}$ with $Y_{\varphi}$ in 
Eqs.~(\ref{evoluzero}), (\ref{Escat}), (\ref{photoc}) and (\ref{Dceq}).
It is again reasonable to assume that the standing wave mode is highly damped,
i.e. $\kappa \gg \chi$, so that it is possible to eliminate it adiabatically.
The perturbative expansion (\ref{Dofrho}) now becomes
\begin{eqnarray}\label{Dofrho2}
 D_c&=&\left(\rho_c-\frac{\chi^2}{\kappa ^2}X\rho_c X\right)
\otimes|0\rangle\langle 0|
-i\frac{\chi}{\kappa}\left(
X\rho_c\otimes|1\rangle\langle 0|
-\rho_c X\otimes|0\rangle\langle 1|\right)\nonumber \\
&+& \frac{\chi^2}{\kappa ^2}X\rho_c X \otimes |1\rangle \langle 1|
-  \frac{\chi^2}{\kappa^2 \sqrt{2}}\left(
X^2\rho_c\otimes|2\rangle\langle 0|
+\rho_c X^2\otimes|0\rangle\langle 2|\right) \,,
\end{eqnarray}
where $|n\rangle $, $n=0,1,2$, are the lowest standing wave mode Fock states.
Using this adiabatic expansion and tracing over the standing wave mode, 
one finally gets exactly 
Eq.~(\ref{rhoceq}), describing the reduced dynamics of the vibrational
mode conditioned to the result $\langle X(t)\rangle_c$ of the continuous 
position measurement.

\section{The Feedback Loop}

We are now able to use the continuous record of the atom position
to control its motion through the application of a feedback loop. 
We shall use the continous 
feedback theory proposed by Wiseman and Milburn 
\cite{wisemil}.

One has to take part 
of the stochastic output homodyne 
photocurrent $I(t)$, obtained from the continuous monitoring 
of the meter mode, and 
feed it back to the vibrational dynamics 
(for example as a driving term)
in order to modify the evolution of the mode $a$. 
To be more specific, the presence of feedback modifies the 
evolution of the conditioned state
$\rho_c(t)$. It is reasonable to assume that the feedback 
effect can be described by an additional
term in the master equation, linear in the photocurrent 
$I(t)$, i.e. \cite{wisemil}
\begin{equation}\label{rhofb}
\left[{\dot\rho}_c(t)\right]_{fb}=\frac{I(t-\tau)}{\eta\chi}\,
{\cal K}\rho_c(t)\,,
\end{equation}
where $\tau$ is the time delay in the feedback loop
and ${\cal K}$ 
is a Liouville superoperator describing the way 
in which the feedback signal acts 
on the system of interest.  

The feedback term (\ref{rhofb}) has to be considered in the 
Stratonovich sense, since Eq.
(\ref{rhofb}) is introduced as limit of a real process, then 
it should be transformed in the Ito
sense and added to the evolution equation (\ref{rhoceq}). 
A successive average over the white noise
$\xi(t)$ yields the master equation  for the vibrational 
density matrix $\rho$ 
in the presence of feedback. 
Only in the limiting case of a feedback
delay time much shorter than the characteristic time of the 
$a$ mode, it is possible to obtain a Markovian 
equation \cite{wisemil,delay}, which is given by
\begin{equation}
\dot{\rho }={\cal L}_{h} \rho
-i\nu \left[a^{\dagger}a,\rho\right] -\frac{\chi^2}{2\kappa}\left[X,\left[X 
,\rho \right]\right]+{\cal K}\left(ie^{i\varphi }\rho X-
ie^{-i\varphi} X\rho\right)
+\frac{{\cal K}^{2}}{2\eta\chi^2/\kappa}\rho.
\label{qndfgen}
\end{equation}
The second term of the right hand side of Eq.~(\ref{qndfgen}) 
is the 
usual double-commutator term associated to the 
measurement of 
$X$; the third term is the feedback term itself and 
the fourth 
term is a diffusion-like term, which is an unavoidable 
consequence of the noise introduced 
by the feedback itself.

Then, 
since the Liouville superoperator ${\cal K}$ can only 
be of Hamiltonian 
form \cite{wisemil}, we choose it as 
${\cal K}\rho =-ig 
\left[P,
\rho \right]/2$ \cite{TV}, where $P=(a-a^{\dagger})/2i$ is
the adimensional momentum operator of the trapped particle
and
$g$ is the feedback gain related to the 
practical way of realizing the loop. 
One could have chosen to feed
the system with a generic phase-dependent quadrature; 
however, it is possible to see
that the above choice gives the best and simplest result \cite{TV}.
Using the above expressions in
Eq.~(\ref{qndfgen}) and rearranging the terms in 
an appropriate 
way, we finally get the following master equation:
\begin{eqnarray}\label{totale}
\dot{\rho }&=& \frac{\Gamma}{2}(N+1)
\left(2a\rho a^{\dagger}-a^{\dagger}a\rho 
-\rho a^{\dagger}a\right)
+\frac{\Gamma}{2}N
\left(2a^{\dagger}\rho a-aa^{\dagger}\rho 
-\rho aa^{\dagger}\right)
\nonumber \\
&-&\frac{\Gamma}{2}M
\left(2a^{\dagger}\rho a^{\dagger}-a^{\dagger 2}\rho 
-\rho a^{\dagger 2}
\right)
-\frac{\Gamma}{2}M^{*}
\left(2a\rho a-a^{2}\rho -\rho a^{2}\right)
\nonumber \\
&-&\frac{g}{4}\sin\varphi
\left[a^{2}-a^{\dag 2},\rho\right]-i\nu\left[a^{\dag}a,\rho\right]\,,
\end{eqnarray}
where 
\begin{eqnarray}\label{parameters}
\Gamma&=&-g\sin\varphi\,;\\
N&=&-\frac{1}{g\sin\varphi }\left[\gamma_h
+\frac{\chi^2 
}{4\kappa}+\frac{g^{2}}{4\eta\chi^2/\kappa}\right]-\frac{1}{2} \,;\\
M&=&\frac{1}{g\sin\varphi}\left[
\frac{\chi^2}{4\kappa}-\frac{g^{2}}{4\eta\chi^2/\kappa}\right]
-\frac{i}{2}\cot\varphi\,.
\end{eqnarray}
Eq.~(\ref{totale}) is very instructive because 
it clearly shows the 
effects of the feedback loop on the 
vibrational mode $a$. 
The proposed
feedback mechanism, indeed, 
not only introduces a parametric driving term proportional to
$g\sin \varphi$, but it also
simulates the presence of a squeezed bath \cite{qnoise}, 
characterized by an effective damping 
constant $\Gamma $ and by the coefficients
$M$ and $N$, which are given 
in terms of the feedback parameters \cite{TV}. 
An interesting aspect of the effective bath described by 
the first four 
terms in the right hand side of (\ref{totale}) is that 
it is characterized by 
phase-sensitive fluctuations, depending upon the 
experimentally adjustable 
phase $\varphi$. This master equation preserves the positivity of
$\rho$ provided that the condition $|M|^2 \leq N(N+1)$ is satisfied
\cite{qnoise}, as it can be checked in the present case.
In fact, under this condition it is always possible 
to find a unitary transformation transforming Eq.~(\ref{totale})
into a master equation manifestly of the Lindblad form.

\section{The Stationary Solution}

Because of its linearity, the solution of  
Eq. (\ref{totale}) can be easily obtained
by using the normally ordered 
characteristic 
function \cite{qnoise} ${\cal C}(\lambda,\lambda^*,t)$.
The partial differential equation corresponding to Eq. 
(\ref{totale}) is
\begin{eqnarray}\label{chareq}
&&\left\{\partial_t+\left(\frac{\Gamma}{2}+i\nu\right)\lambda\partial_{\lambda}
+\left(\frac{\Gamma}{2}-i\nu\right)\lambda^*\partial_{\lambda^*}
+\frac{g}{2}\sin\varphi\left(\lambda\partial_{\lambda^*}
+\lambda^*\partial_{\lambda}\right)\right\}{\cal C}(\lambda,
\lambda^*,t)\nonumber\\
&&=\left\{-\Gamma N|\lambda|^2+
\left(\frac{\Gamma}{2}M+\frac{g}{4}\sin\varphi\right)(\lambda^*)^2
+\left(\frac{\Gamma}{2}M^*+\frac{g}{4}\sin\varphi\right)
\lambda^2\right\}
{\cal C}(\lambda,\lambda^*,t)\,,
\end{eqnarray}

The stationary state is reached only if the parameters 
satisfy the stability condition
that all the eigenvalues have positive real part, which in the present
case is achieved when $g \sin\varphi <0$. In this case
the stationary solution has the following form
\begin{equation}\label{charsol}
{\cal C}(\lambda,\lambda^*,\infty)
=\exp\left[-\zeta|\lambda|^2+\frac{1}{2}\mu(\lambda^*)^2
+\frac{1}{2}\mu^*\lambda^2\right]\,,
\end{equation}
where
\begin{eqnarray}
\zeta&=&\frac{N(g^2\sin^2\varphi+4\nu^2)+g\sin\varphi
\left(2\nu{\rm Im}\{M\}-g\sin\varphi {\rm Re}\{M\}\right)
+g^2\sin^2\varphi/2}
{4\nu^2}\,;\label{ze}\\
\mu&=&\Gamma\frac{(N+1/2)g\sin\varphi+\Gamma {\rm Re}\{M\}+2\nu{\rm Im}\{M\}}
{4\nu^2}\nonumber\\
&+&i\frac{g\sin\varphi}{2\nu}\left[{\rm Re}\{M\}-(N+1/2)\right]
\label{mu}\,.
\end{eqnarray}
This means that in general, the stationary state is a generalized 
gaussian state. However, in practical
situations, the stationary state assumes a much simpler form.
In fact, the vibrational frequency $\nu$ is usually much larger than
the heating rate $\gamma_{h}$ and the feedback parameter $g$, 
and in this limit, Eqs.~(\ref{ze}) 
and (\ref{mu}) become
$\zeta\approx N$ and $\mu\approx 0$ respectively.
Using Eq.~(\ref{charsol}), one has
\begin{equation}\label{charsol2}
{\cal C}(\lambda,\lambda^*,\infty)
=\exp\left[-N|\lambda|^2 \right]\,,
\end{equation}
that is, the stationary state is an effective thermal state
with mean vibrational number $N$. 
This can also be seen from the fact that in the large $\nu$ limit,
one can consider the master equation (\ref{totale}) in the frame rotating at 
the frequency $\nu$, and neglecting the rapidly oscillating terms, one ends up
with a thermal master equation given by the first line of 
Eq.~(\ref{totale}), whose steady state is just the thermal state with mean
phonon number $N$.

The expression for $N$ given by Eq.~(\ref{parameters}) shows
that it is convenient to choose $\varphi =-\pi/2$ to get
the smallest possible values for $N$. In this way the stability 
condition is also automatically satisfied. Then, the minimum value for 
the stationary mean vibrational number $N$ can be obtained by 
minimizing it with respect to the feedback gain $g$: the optimal value 
for $g$ is given by $g=4\left(\left(\gamma_{h}+\chi^{2}/4\kappa\right)
\eta \chi^{2}/4\kappa\right)^{1/2}$ and the corresponding minimum 
value of $N$ is
\begin{equation}
N_{min} = \frac{1}{2}\left[\sqrt{\frac{1+4\kappa 
\gamma_{h}/\chi^{2}}{\eta}}-1\right]\;.
\label{nmin}
\end{equation}
This expression shows the best cooling result achievable with the 
present feedback scheme. One has that when the heating 
rate is negligible with respect to the ``QND coupling'' parameter 
$\chi^{2}/4 \kappa $, the final vibrational number $N$ is limited only by
the efficiency of the homodyne measurement $\eta $. 
In particular this fact is true in the optimal case where all the technical
sources of heating are eliminated which means $\gamma_h=0$ in all the
above expressions. Therefore, 
the present scheme is able to achieve cooling to the motional ground
state especially in the off-resonant scheme, in which the particle 
position measurement is realized through homodyning the radiation 
exiting the cavity. In fact, in this case the homodyne efficiency can be 
very close to one. This is shown in Fig.~1, where
we have sketched the phase space uncertainty 
contours obtained by cutting the
Wigner function corresponding to Eq.(\ref{charsol}) at 
$1/\sqrt{e}$ times its maximum height. 
We see that the feedback produces a relevant contraction of 
the uncertainty region, which becomes
almost indistinguishable from the region 
corresponding to the motional ground state (inner dotted line in Fig.~1).
The outer dashed line corresponds instead to the initial thermal state
with mean vibrational number $n_{0}=10$, prepared by the Doppler 
pre-cooling stage.

In the resonant case in which the effective position measurement
is realized through the homodyne measurement of the fluorescence, the
measurement efficiency is much lower and ground state cooling becomes
very difficult to achieve. However, this
position measurement scheme based on fluorescence becomes necessary when one 
cannot extract the light out of a cavity, such as in Ref. \cite{hof},
and one has to use the fluorescent light.

\section{Conclusions}

We have proposed a feedback scheme able to achieve significant cooling
of the motional degree of freedom of a trapped particle. The method is
based on an effective continuous measurement of the particle position
which can be realized in two different ways: by homodyning either
the fluorescence of a strong transition or directly the light exiting 
the cavity which is coupled to the trapped particle.
When the efficiency of the homodyne detection is close to one,
the method is able to achieve ground state cooling.
It is interesting to note that, even if only the particle position
is measured and the feedback is chosen in order to decrease position 
fluctuations, the scheme provides a
{\em phase-space} localisation for all quadratures.
This is
essentially due to the fact that the bare atom Hamiltonian $\hbar \nu 
a^{\dagger}a$ mixes the dynamics of the atomic position and momentum, so
that the continuous homodyne measurement actually gives
informations on both quadratures.
This model shares some peculiarities with
that one we have proposed in \cite{prl}
to cool the vibrational motion of a macroscopic mirror of an optical
cavity. 
The present application to a trapped atom stresses the versatility of 
the methods using feedback loops in systems characterized by the
radiation pressure force in controlling thermal noise.

It is also possible to see that the proposed scheme is not
able to reduce the noise below the quantum limit, i.e. 
the stationary state variance of a generic motional quadrature 
cannot be made smaller than $1/4$.
This is again a consequence
of the free vibrational Hamiltonian $\hbar \nu 
a^{\dagger}a$ (it is in fact possible to get position squeezing in the
limiting case $\nu =0$ \cite{my}).
Actually, position squeezing can be obtained by considering a suitable 
modification of the present scheme \cite{squa}.

As concerns the specific way
in which a particular feedback
Hamiltonian could be implemented, the important
point is to be able to realize a term in 
the feedback Hamiltonian
proportional to momentum. 
This is not straightforward, but could be realized by using the 
feedback current to vary
an external potential applied to the atom without altering 
the trapping potential.
On the other hand, shifts in the position
(being strictly equivalent to a linear momentum term in the Hamiltonian),
are achieved simply by shifting all the position dependent terms in the
Hamiltonian, in particular the trapping potential. 
Alternatively, the use of laser pulses could be useful as well,
since, using a typical laser cooling scheme, the light can exert
on the atom a force proportional to its momentum. 

As we have already remarked, 
in principle the model could be extended to the three 
dimensional case.
As concerns the model discussed in Sec. \ref{sub1},
one should consider three different internal transitions,
each one coupled with a vibrational degree of freedom, resonant with 
three orthogonal standing waves.
For the off-resonant case presented in Sec. \ref{sub2},
one should only consider three orthogonal standing waves
far from resonant transitions.

In conclusion, although the implementation of the 
presented cooling method via feedback could be an hard task,  
it can be useful whenever the use of resolved sideband
cooling is impractical. The possibility of having
an alternative way to cool trapped particles is particularly
interesting for quantum information processing applications,
because it may happen that the requirement of having two 
highly stable internal states for quantum logic operations and
a good internal transition for sideband cooling cannot be
simultaneously satisfied. With this respect other cooling strategies have been
recently proposed, as for example the use of sympathetic cooling
between two different species of ions \cite{peik,wizu}.

\section*{Acknowledgements}

This work has been partially supported by INFM (through the
Advanced Research Project ``CAT"),
by the European Union in the framework of the TMR Network
``Microlasers and Cavity QED".

\bibliographystyle{unsrt}

\begin{thebibliography}{99}

\bibitem{pg}
see e.g., P. K. Ghosh, {\it Ion Traps}, (Clarendon, Oxford, 1995), 
and references therein.

\bibitem{wineprl}
F. Diedrich, J.C. Bergquist, W.M. Itano and D.J. Wineland, 
Phys. Rev. Lett. {\bf 62}, 403 (1989).

\bibitem{optlat}
S.E. Hamann, D.L. Haycock, G. Close, P.H. Pax,
I.H. Deutsch, and P.S. Jessen, Phys. Rev. Lett. {\bf 80}, 4149 (1998);
M.T. DePue, C. McCormick, S.L. Winoto, S. Oliver, and D.W. Weiss, 
Phys. Rev. Lett. {\bf 82}, 2262 (1999);
S. Friebel, C. D'Andrea, J. Walz, M. Weitz, and T.W. Hansch,
Phys. Rev. A {\bf 57}, R20 (1998).

\bibitem{ciraczoller} 
J.I. Cirac, P. Zoller, Phys. Rev. Lett. {\bf 74}, 4091 (1995).

\bibitem{peik} E. Peik, J. Abel, Th. Becker, J. von Zanthier, and
H. Walther, Phys. Rev. A {\bf 60}, 439 (1999).

\bibitem{monroe}C. Monroe, D.M. Meekhof, B.E. King, S.R. Jefferts,
W.M. Itano, D.J. Wineland, and P. Gould, Phys. Rev. Lett. {\bf 75}, 4011 
(1995).

\bibitem{nist}
D.J. Wineland, C. Monroe, W.M. Itano, 
D. Leibfried, B.E. King,
D.M. Meekhof, J. Res. Natl. Inst. Stand. Technol. {\bf 103}, 259 (1998).

\bibitem{sara}
T.A. Savard, K.M. O'Hara, and J.E. Thomas, Phys. Rev. A {\bf 56},
R1095 (1997).

\bibitem{qo}
see e.g., D. F. Walls and G. J. Milburn,
{\it Quantum Optics}, (Springer, Berlin, 1994).

\bibitem{danrisk}
C.A. Blockley, D. F. Walls and H. Risken, 
Europhys. Lett. {\bf 17}, 509 (1992).

\bibitem{cirac}
J.I. Cirac, R. Blatt, A. S. Parkins and P. Zoller, 
Phys. Rev. A {\bf 49}, 1202 (1994).

\bibitem{sara2}S. Schneider, G.J. Milburn, Phys. Rev. A {\bf 59}, 3766 (1999).

\bibitem{hof}
J. T. H\"offges, H. W. Baldauf, T. Eichler, S. R. Helmfrid and H. Walther, Opt. Comm.
{\bf 133}, 177 (1997).

\bibitem{vog}
W. Vogel, Phys. Rev. Lett. {\bf 67}, 2450 (1991);
Phys. Rev. A {\bf 51}, 4160 (1995).


\bibitem{quadra}
H.M. Wiseman, and 
G.J. Milburn, Phys. Rev. A {\bf 47}, 
642 (1993).

\bibitem{TV}
P. Tombesi and D. Vitali, 
Appl. Phys. B {\bf 60}, S69 (1995);
Phys. Rev. A {\bf 51}, 4913 (1995).


\bibitem{wisemil}
H.M. Wiseman and G.J. Milburn,
Phys. Rev. Lett. {\bf 70}, 
548 (1993); Phys. Rev. A {\bf 49}, 1350 (1994);
H.M. Wiseman, Phys. Rev. A {\bf 49}, 2133 (1994).

\bibitem{delay}
It has been recently shown (V. Giovannetti, P. Tombesi and D.
Vitali, Phys. Rev. A, {\bf 60}, 1549 (1999)) 
that it is actually possible to solve the general
non-Markovian problem in the presence of a nonzero delay. 

\bibitem{qnoise}
C.W. Gardiner, {\em Quantum Noise} 
(Springer, Berlin, 1991).

\bibitem{prl}
S. Mancini, D. Vitali, and P. Tombesi, Phys. Rev. Lett.
{\bf 80}, 688 (1998);
this model has been experimentally implemented by
P. F. Cohadon, A. Heidmann and M. Pinard,
Phys. Rev. Lett. {\bf 83}, 3174 (1999).


\bibitem{my} Some preliminary results are given in S. Mancini, Acta Phys. 
Slovaca {\bf 49}, 725 (1999).

\bibitem{squa}S. Mancini, D. Vitali, and P. Tombesi, in press on 
J. Opt. B: Quantum. Semiclass. Opt.

\bibitem{wizu}D. Kielpinski {\it et al.}, quant-ph/9909035.


\end{thebibliography}


\begin{figure}[htb]

\centerline{\epsfig{figure=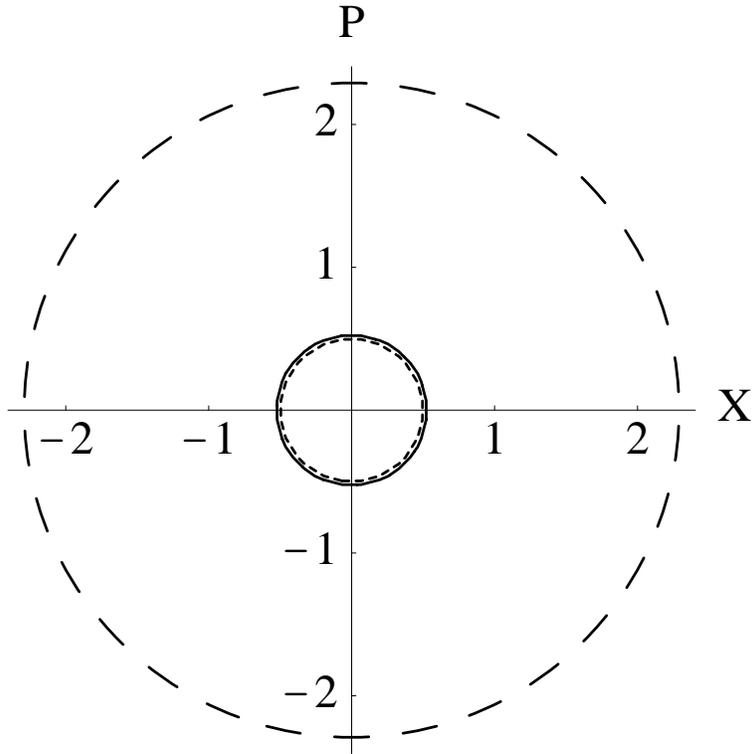,height=10cm}}
\caption{Phase space uncertainty 
contours obtained by cutting the
Wigner function of the stationary state at 
$1/\sqrt{e}$ times its maximum height.
The dashed line refers to the initial thermal state with mean
vibrational number $n_0=10$; the solid line refers to the steady state
in the presence of feedback with
$\chi=4$ kHz, $g=0.375$ kHz, 
$\nu=1$ MHz,
$\gamma_h=10$ Hz, $\kappa=40$ kHz,
$\eta=0.9$, $\varphi=-\pi/2$.
Notice that feedback provides cooling very close to the ground state
(the corresponding contour is given by the dotted line).}
\label{fig1}

\end{figure}

\end{document}